\documentclass[aps,prd,twocolumn,nofootinbib]{revtex4}

\usepackage{float}
\usepackage{booktabs,makecell,multirow}
\usepackage{longtable}
\usepackage{graphicx}
\usepackage{color}
\usepackage[colorlinks=true, linkcolor=blue, citecolor=blue, urlcolor=blue]{hyperref}

\newcommand*{\RMN}[1]{\uppercase\expandafter{\romannumeral#1}}

\begin{document}

\title{New determination of $|V_{\rm cb}|$ using the three-loop QCD corrections for the $B\to D^{\ast}$ semi-leptonic decays}

\author{Hua Zhou$^{1,2}$}
\email{zhouhua@cqu.edu.cn}

\author{Qing Yu$^{1,2}$}
\email{yuq@cqu.edu.cn}

\author{Xu-Chang Zheng$^{1}$}
\email{zhengxc@cqu.edu.cn}

\author{Hai-Bing Fu$^3$}
\email{fuhb@cqu.edu.cn}

\author{Xing-Gang Wu$^{1}$}
\email{wuxg@cqu.edu.cn}

\affiliation{$^1$Department of Physics, Chongqing Key Laboratory for Strongly Coupled Physics, Chongqing University, Chongqing 401331, P.R. China \\
$^2$ Department of Physics, Norwegian University of Science and Technology, H{\o}gskoleringen 5, N-7491 Trondheim, Norway \\
$^3$Department of Physics, Guizhou Minzu University, Guiyang 550025, P.R. China}

\date{\today}

\begin{abstract}

We present a new determination of the Cabibbo-Kobayashi-Maskawa matrix element $|V_{\rm cb}|$ by using the three-loop perturbative QCD corrections for the $B\to D^{\ast}$ semi-leptonic decay. The decay width of $B\to D^{\ast}$ semi-leptonic decay can be factorized as perturbatively calculable short-distance part and the non-perturbative but universal long-distance part. We adopt the principle of maximum conformality (PMC) single-scale setting approach to deal with the perturbative series so as to achieve a precise fixed-order prediction for the short-distance parameter $\eta_{A}$. By applying the PMC, an overall effective $\alpha_s$ value is achieved by recursively using the renormalization group equation, which inversely results in a precise scale-invariant pQCD series. Such scale-invariant series also provides a reliable basis for predicting the contributions from uncalculated perturbative terms. We then obtain $\eta_{A}=0.9225^{+0.0117}_{-0.0168}$, where the error is the squared average of those from $\Delta\alpha_{s}(M_Z)=\pm0.0010$ and the uncertainties caused by the uncalculated higher-order perturbative terms. By using the data of $B\to D^{\ast}\ell\bar{\nu}_{\ell}$, we finally obtain $|V_{\rm cb}|_{\rm PMC} =(40.60^{+0.53}_{-0.57})\times10^{-3}$, which is consistent with the PDG value within errors.

\end{abstract}

\maketitle

The $|V_{\rm cb}|$ is an important element of the Cabbibo-Kobayashi-Maskawa (CKM) matrix, whose precise value is helpful for precision test of the Standard Model. Due to the recent theoretical progresses and a plentiful collection of the data on the $B$-meson semileptonic decays at the $B$ factories and the large hadronic collider (LHC), we are facing the chance of achieving more precise CKM matrix elements. Among them, the $B$-meson decays to charmed $D^{\ast}$-meson, $B\to D^{\ast} \ell \bar\nu_{\ell}$ with $\ell$ being the light leptons, is helpful for extracting $|V_{\rm cb}|$~\cite{Isgur:1989vq, Boyd:1997kz, Caprini:1997mu, Abbiendi:2000hk, Abreu:2001ic, Abdallah:2004rz, Aubert:2007rs, Aubert:2007qs, Aubert:2008yv, Waheed:2018djm, FermilabLattice:2014ysv, VaqueroAviles-Casco:2017cjb, Ferlewicz:2020lxm, Ricciardi:2019zph, Martinelli:2021und}. Using the heavy-quark symmetry, it's differential decay width $d\Gamma/dw$ can be written as~\cite{Neubert:1994vy}
\begin{eqnarray}
\frac{d\Gamma(B\rightarrow D^{\ast}{\ell}\bar{\nu}_{\ell})}{dw}&=&\frac{G^{2}_{F}}{48\pi^{3}}(m_{B}-m_{D^{\ast}})^{2} \sqrt{\omega^{2}-1}(\omega+1)^{2}\nonumber\\&&
\left[1+\frac{4\omega}{\omega+1}\frac{m^{2}_{B}-2\omega m_{B} m_{D^{\ast}}+m^{2}_{D^{\ast}}}{(m_{B}-m_{D^{\ast}})^{2}}\right]\nonumber\\
&& \times m^{3}_{D^{\ast}} |V_{\rm cb}|^{2} {\cal F}^{2}(\omega),
\label{Diff-width}
\end{eqnarray}
where ${\cal F}(\omega)$ is the hadronic form factor, $G_{F}$ is the Fermi constant, $m_{B}$ and $m_{D^{\ast}}$ are $B$-meson and $D^{\ast}$-meson masses, respectively. The kinematic variable $w=v\cdot v^{\prime}={(m_B^2+m_{D^{\ast}}^2-q^2)}/{2m_B m_{D^{\ast}}}$ represents the four-velocity transfer between the $B$-meson velocity $v$ and the $D^{\ast}$-meson velocity $v^\prime$, where $q^2=(p_B - p_{D^{\ast}})^2$.

Using the experimental data on the differential decay width $d\Gamma/dw$, one can fix the value of the combined parameter $|V_{\rm cb}|{\cal F}(\omega)$ at various momentum transfers. Then, if one has known precise value of the hadronic form factor ${\cal F}(\omega)$, one can finally achieve precise value of $|V_{\rm cb}|$. At present, ${\cal F}(\omega)$ has been calculated under various approaches, such as the Lattice QCD~\cite{FermilabLattice:2014ysv, VaqueroAviles-Casco:2017cjb, Ferlewicz:2020lxm}, the QCD Sum Rule~\cite{Shifman:1994jh}, the covariant light-front quark model~\cite{Kang:2018jzg, Zhang:2020dla} and the Perturbation QCD (pQCD)~\cite{Czarnecki:1996gu, Archambault:2004zs}, and etc. Using the pQCD factorization approach, ${\cal F}(\omega)$ can be expressed as a product of the short-distance coefficient $\eta_{A}$ and the long-distance hadronic dynamics $\hat{\xi}(\omega)$, e.g. ${\cal F}(\omega)=\eta_{A}\hat{\xi}(\omega)$. Experimentally, one usually extracts the magnitude of $|V_{\rm cb}|$ by using the product $|V_{\rm cb}|{\cal F}(\omega)$ at the zero recoil point $w=1$ (or $q^2=0$)~\cite{Abbiendi:2000hk, Waheed:2018djm, Aubert:2008yv, Aubert:2007rs, Aubert:2007qs, Abreu:2001ic, Abdallah:2004rz}. At this point, we have $\hat{\xi}(1)=1+{\cal O}({1/m^{i}_Q})$ by using the heavy quark effective theory~\cite{Luke:1990eg}, where $Q$ stands for $m_c$ or $m_b$ and the power correction have been considered in refs~\cite{Neubert:1995bc, Gambino:2010bp, Gambino:2012rd}, at the ${\cal O}({1/m^{2}_Q})$-order level is about $-(5.5\pm2.5)\%$. Thus the key component of improving the accuracy of the differential decay width is to achieve a precise prediction on the short-distance parameter $\eta_{A}$ at the zero recoil point, whose perturbative expressions can be expressed as
\begin{eqnarray}
\eta_{A} &=& 1+C_{F} \sum_{i=1}^{\infty} \eta^{(i)}_{A} a^{i}_{s}(\mu_r), \label{pQCDetaA}
\end{eqnarray}
where $a_s=\alpha_s/\pi$, $C_{F}={N^{2}_{c}-1}/{(2N_{c})}$ for ${\rm SU}(N_{c})$ color group, and $\mu_r$ is the renormalization scale. The perturbative coefficients $\eta^{(i)}_{A}$ has been calculated up to three-loop QCD corrections under the conventional $\overline{\rm MS}$-scheme~\cite{Archambault:2004zs}.

It has been found that the pQCD series (\ref{pQCDetaA}) involves two mass scales $m_b$ and $m_c$, one usually sets the renormalization scale $\mu_r$ as the typical scale $Q=\sqrt{m_b m_c}$ and vary it within a certain range such as $[\sqrt{m_b m_c}/2, 2\sqrt{m_b m_c}]$ to ascertain its uncertainty. However as shall be shown below, large scale uncertainties persist for such simple treatment even when more loop terms have been included. This is due to the divergent renormalon terms such as $n ! \beta^n_0 \alpha_s^n$ (The $\beta_i$-functions satisfy the approximation $\beta_i \approx \beta_0^{i+1}$)~\cite{Beneke:1994qe, Neubert:1994vb, Beneke:1998ui}, and the mismatching of the $\alpha_s$ and its perturbative coefficients. A valid pQCD prediction for a physical observable should be independent to any choices of renormalization scheme and renormalization scale. However a truncated perturbation series does not automatically satisfy these requirements. Especially, by using the guessed scale, it will generally violate the renormalization group invariance of the pQCD approximant~\cite{Wu:2013ei, Wu:2019mky, Wu:2014iba} and then decreases the predictive power of perturbative calculation. The error estimate obtained by varying the scale within an ``ad hoc" range can obtain some information from the $\beta$-dependent terms, but not from the conformal terms at the higher-orders. Moreover, the large scale dependence of each loop terms also makes it hardly be a reliable basis to estimate the contributions from uncalculated terms. Thus to improve the accuracy of pQCD prediction, it is necessary to eliminate such scale ambiguity. In this article, we will use the principle of maximum conformality (PMC)~\cite{Brodsky:2011ta, Brodsky:2011ig, Brodsky:2012rj, Mojaza:2012mf, Brodsky:2013vpa} to reanalyze this process, showing how the $|V_{\rm cb}|$ could be improved when the conventional scale-setting ambiguity has been removed from the parameter $\eta_A$. The PMC extends the Brodsky-Lepage-Mackenzie approach~\cite{Brodsky:1982gc} for scale-setting in pQCD to all orders, and it provides a systematical way to eliminate conventional scale-setting ambiguity by recursively using the renormalization group equation (RGE). The RGE is applied for determining the correct $\alpha_s$ running behavior by using the $\{\beta_i\}$-terms of the perturbative series; and as a byproduct, the perturbative convergence of the series can be generally improved due to the elimination of the RGE-involved renormalon terms. It has been demonstrated that after applying the PMC, the perturbative series becomes scale-invariant series independent to any choice of $\mu_r$~\cite{Wu:2018cmb}. Using the known perturbative series (\ref{pQCDetaA}) under the $\overline{\rm MS}$-scheme up to three-loop QCD corrections \cite{Archambault:2004zs}, we have
\begin{eqnarray}
\eta_{A} &=& 1 - 0.667 a^{\overline{\rm MS}}_{s}(\mu_r) + \Bigg[ -2.501+ 0.130n_f \nonumber\\
&&\!\!\!\!  + 0.349n_{f}\log\left(\frac{\mu^{2}_r}{Q^{2}}\right) -5.760\log\left(\frac{\mu^{2}_r}{Q^{2}}\right)  \Bigg]a^{\overline{\rm MS}, 2}_{s}(\mu_r) \nonumber\\
 && \!\!\!\!\!\!\!\! + \Bigg[-32.612+4.782n_{f} -0.096n^{2}_{f} -85.157\log\left(\frac{\mu^{2}_r}{Q^{2}}\right) \nonumber\\
 && +6.031n_{f}\log^{2}\left(\frac{\mu^{2}_r}{Q^{2}}\right) +10.068n_{f}\log\left(\frac{\mu^{2}_r}{Q^{2}}\right) \nonumber\\
 && -0.183n^{2}_{f} \log^{2}\left(\frac{\mu^{2}_r}{Q^{2}}\right) - 0.136n^{2}_{f}\log\left(\frac{\mu^{2}_r}{Q^{2}}\right) \nonumber\\
 && -49.759\log^{2}\left(\frac{\mu^{2}_r}{Q^{2}}\right)\Bigg]a^{\overline{\rm MS}, 3}_{s}(\mu_r) + {\cal O}(a^{\overline{\rm MS}, 4}_{s}),
\label{tab333}
\end{eqnarray}
where $Q=\sqrt{m_{c} m_{b}}$. It should be emphasized that the PMC method determines the correct momentum flow by absorbing the non-conformal $\{\beta_i\}$-terms of the perturbation sequence. Therefore, in order to correctly distinguish the conformal/non-conformal terms, the $n_l$ in the perturbation sequence is converted to the number of active light quark flavors $n_f$, and $n_f=n_l+n_H$. Here $n_l=3$ denotes the number of massless quarks and $n_H=2$. The pole masses $m_{b}=4.78~\rm GeV$ and $m_{c}=1.68~\rm GeV$ are implicitly adopted in those coefficients~\footnote{Here $m_b$ and $m_c$ are pole quark masses. Since when using the PMC, it is helpful to use the pole mass other than the $\overline{\rm MS}$-mass in the perturbative series so as to find the correct $\{\beta_i\}$-terms for fixing the $\alpha_s$ running behavior.}.

After applying the PMC, all the scheme-dependent RGE-involved non-conformal $\{\beta_i\}$-terms have been removed, the resultant series becomes scheme-independent and the conventional renormalization scheme ambiguity can also be eliminated. This scheme independence can also be ensured by the commensurate scale relations among the pQCD approximants under various schemes~\cite{Brodsky:1994eh, Huang:2020gic}. Though scheme independent, sometimes, a proper choice of scheme is helpful to improve the pQCD convergence and to avoid small scale problem and achieve a reliable pQCD prediction~\footnote{Practically, the determined PMC scale may be close to or less than the critical scale $\Lambda_{\rm QCD}$ under certain scheme, and a low-energy $\alpha_s$ model has to be chosen to get a reasonable prediction.}. For the present case, to avoid the small scale problem, we transform the above $\overline{\rm MS}$-scheme series into the $V$-scheme series~\cite{Appelquist:1977tw, Fischler:1977yf, Peter:1996ig, Schroder:1998vy, Yu:2021yvw}. The $V$-scheme is a physical scheme, which sums up the effects of gluon exchanges at the low momentum transfer, corrects the static potential by including higher-order QCD corrections, and is gauge-invariant. In the literature, the $V$-scheme has been successfully applied in various phenomenologically oriented QCD studies, such as the heavy-quark production~\cite{Brodsky:1995ds}, the hard-scattering matrix elements of exclusive processes~\cite{Brodsky:1997dh}, and to achieve a smooth transition through the thresholds of heavy quark productions~\cite{Brodsky:1998mf}. To do the transformation, we adopt the following relation between the couplings under the $\overline{\rm MS}$-scheme and $V$-scheme:
\begin{eqnarray}
a_s^{\overline{\rm{MS}}}(\mu_r) = a_s^{\rm{V}}(\mu_r)\bigg[1+\sum_{i=1}^{\infty} y_{i} a_s^{\rm{V}, i}(\mu_r)\bigg],
\end{eqnarray}
whose first two coefficients, which are needed for the present three-loop analysis, are~\cite{Kataev:2015yha}
\begin{eqnarray}
y_1&=&-\frac{31}{12}+\frac{5}{18}n_f,\\
y_2&=&-\frac{499}{288}-\frac{9}{4}\pi^2+\frac{9}{64}\pi^4-\frac{33}{8}\zeta_3-\frac{11}{432}n_f \nonumber\\
&& +\frac{13}{12}\zeta_3n_f+\frac{25}{324} n_f^2,
\end{eqnarray}
where $\zeta_3$ is the Riemann zeta function. We are then ready to transform the $\overline{\rm MS}$-scheme perturbative series (\ref{tab333}) into the $V$-scheme one, i.e.
\begin{eqnarray}
\eta_{A} &=& 1+\sum^{3}_{i=1}{r}^{\rm V}_{i} a^{{\rm V}, i}_{s}(\mu_r) + {\cal O}(a^{{\rm V},4}_{s}),
\end{eqnarray}
where
\begin{eqnarray}
r^{\rm V}_{1}&=&-0.667,\\
r^{\rm V}_{2}&=&-0.779-0.056n_{f}+0.349n_{f}\log\left(\frac{\mu^{2}_{r}}{Q^{2}}\right) \nonumber\\
&&
-5.760\log\left(\frac{\mu^{2}_{r}}{Q^{2}}\right), \\
r^{\rm V}_{3}&=&-9.558+1.872n_{f}-0.076n^{2}_{f}+5.064n_{f}\log\left(\frac{\mu^{2}_{r}}{Q^{2}}\right) \nonumber\\
&&
+6.031n_{f}\log^{2}\left(\frac{\mu^{2}_{r}}{Q^{2}}\right) +0.058n^{2}_{f}\log\left(\frac{\mu^{2}_{r}}{Q^{2}}\right) \nonumber\\
&&
-0.183n^{2}_{f}\log^{2}\left(\frac{\mu^{2}_{r}}{Q^{2}}\right) -55.399 \log\left(\frac{\mu^{2}_{r}}{Q^{2}}\right) \nonumber\\
&&
-49.759\log^{2}\left(\frac{\mu^{2}_{r}}{Q^{2}}\right).
\end{eqnarray}
By further using the general QCD degeneracy relations among different orders~\cite{Bi:2015wea}, each perturbative coefficient can be written as a $\{\beta_i\}$-series and we obtain
\begin{eqnarray}
\eta_{A}&=&1+ r^{\rm V}_{1,0}a^{\rm V}_{s}(\mu_r) +(r^{\rm V}_{2,0}+\beta_0 r^{\rm V}_{2,1})a^{{\rm V}, 2}_{s}(\mu_r) \nonumber\\
&+& (r^{\rm V}_{3,0}+\beta_1 r^{\rm V}_{2,1}+2\beta_{0} r^{\rm V}_{3,1}+\beta^2_{0} r^{\rm V}_{3,2}) a^{{\rm V},3}_{s}(\mu_r)\nonumber\\
&+& {\cal O}(a^{{\rm V},4}_{s}),
\label{ag1con}
\end{eqnarray}
where $r^{\rm V}_{i,0}$ are scale-invariant conformal coefficients and $r^{\rm V}_{i,j(\neq0)}$ are generally scale-dependent coefficients, all of which can be derived from the above equations.

Following the standard PMC single-scale setting procedures~\cite{Shen:2017pdu}, all non-conformal $\{\beta_{i}\}$-terms shall be adopted for determining the correct $\alpha_s$-value of the process, and the parameter $\eta_A$ becomes the following perturbative series which is free of RGE-involved $\{\beta_i\}$-terms, i.e.
\begin{eqnarray}
\eta_{A|\rm PMC}&=&1+\sum^{3}_{i=1}{r}^{\rm V}_{i,0} a^{{\rm V},i}_{s}(Q_\ast) + {\cal O}(a^{{\rm V},4}_{s}),
\end{eqnarray}
where $Q_\ast$ represents the correct momentum flow of the process determined by the RGE, and by using the present known three-loop series, it can be fixed up to next-to-leading log (NLL) accuracy,
\begin{eqnarray}
\ln\left(\frac{Q^{2}_{\ast}}{Q^2}\right) &=& T_{0} + T_{1} a^{\rm V}_s(Q) + {\cal O}(a^{\rm V,2}_{s}),
\label{equ1}
\end{eqnarray}
where
\begin{eqnarray}
T_0 &=& -{{r}^{\rm V}_{2,1}\over {r}^{\rm V}_{1,0}}
\end{eqnarray}
and
\begin{eqnarray}
T_1 &=& {2({r}^{\rm V}_{2,0}{r}^{\rm V}_{2,1}-{r}^{\rm V}_{1,0}{r}^{\rm V}_{3,1})\over {r}_{1,0}^{{\rm V} ,2}} +{({r}_{2,1}^{{\rm V},2}-{r}^{\rm V}_{1,0}{r}^{\rm V}_{3,2})\over {r}_{1,0}^{{\rm V},2}}\beta_0.
\end{eqnarray}
It is found that $Q_{\ast}$ is free of the renormalization scale $\mu_{r}$. This indicates that one can take any $\mu_r$ to finish the renormalization procedures for the parameter $\eta_A$, and the resultant PMC series shall be independent to this choice, well satisfying the requirement of RGI.

To do the numerical calculation, we adopt $\alpha_{s}(M_Z)=0.1179\pm0.0010$~\cite{Zyla:2020pa} as the reference point for fixing $\alpha_s$ running behavior, which leads to $\Lambda^{\overline{\rm MS}}_{{\rm QCD}|n_f =5}=207.2^{+11.8}_{-11.4}$ MeV and $\Lambda^{{\rm V}}_{{\rm QCD}|n_f =5}=283.0^{+16.1}_{-15.6}$ MeV, respectively.

\begin{figure}[htb]
\centering
\includegraphics[width=0.48\textwidth]{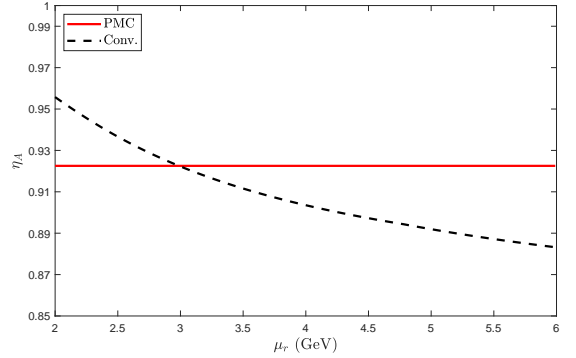}
\caption{The parameter $\eta_{A}$ up to N$^3$LO-level as a function of the renormalization scale $\mu_r$ under conventional and PMC scale-setting approaches. The dashed and the solid lines are for conventional and PMC ones, respectively.}
\label{fig1}
\end{figure}

\begin{table*}[htb]
\begin{tabular}{c c  c  c  c c c}
\hline
& ~$ $~ & ~$\rm LO$~ & ~$\rm NLO$~ & ~$\rm N^{2}LO$~ & ~$\rm N^{3}LO$~ &~$\rm Total$\\
\hline
& $\rm Conv.$ & $1$ & $-0.0625^{-0.0328}_{+0.0147}$ & $-0.0093^{+0.1014}_{-0.0247}$  & $-0.0017^{-0.0242}_{-0.0308}$  & $0.9265^{+0.0444}_{-0.0408}$ \\
& $\rm PMC$ & $1$ & $-0.0644$ & $-0.0158$ & $0.0027$& $0.9225$\\
\hline
\end{tabular}
\caption{The vlaue of $\eta_{A}$ up to $\rm N^{3}LO$ order under conventional and PMC scale-setting approaches. For conventional series, its center value is for $\mu_r=Q$ and its upper and lower errors are for $\mu_r\in[Q/2,2Q]$. }
\label{tab1}
\end{table*}

Firstly, we present the N$^3$LO-$\eta_{A}$ as a function of the renormalization scale $\mu_r$ in Fig.~\ref{fig1}, which are calculated under conventional and PMC scale-setting approaches, respectively. There are large cancellations among different orders for conventional series, leading to a small scale error for the N$^3$LO-level prediction; e.g. by taking $\mu_{r}\in[Q/{2}, 2Q]$, the net scale error of $\eta_A$ is
\begin{equation}
\Delta\eta_A|_{\rm Conv.}=\left(^{+0.0444}_{-0.0408}\right).
\end{equation}
This is clearly shown in Table~\ref{tab1}, which indicates that the scale errors of each loop terms are rather large. While the PMC predictions are independent to any choice of $\mu_r$ for either each loop terms or the total series.

\begin{table}[htb]
\begin{tabular}{c c c  c}
\hline
 ~ ~ & ~$k_1$~ & ~$k_2$~ & ~$k_3$\\
\hline
 $\rm Conv.$ & $-6.25\%^{-3.28\%}_{+1.47\%}$ & $-0.93\%^{+10.14\%}_{-2.47\%}$ & $-0.17\%^{-2.42\%}_{-3.08\%}$ \\
 $\rm PMC$  & $-6.44\%$& $-1.58\%$ & $0.27\%$ \\
\hline
\end{tabular}
\caption{The ratio $k_i$ of the perturbation series under conventional and PMC scale-setting approaches, respectively.}
\label{tab2}
\end{table}

We define a ratio $k_i$ to show the convergent behavior of the perturbative series, e.g.,
\begin{eqnarray}
k_i={\Gamma^{\rm N^{i}LO}}/{\Gamma^{\rm LO}}.
\end{eqnarray}
Table~\ref{tab2} shows the ratios $k_i$ up to N$^3$LO-level under conventional and PMC scale-setting approaches, respectively. The PMC series is scale-invariant and convergent, and its ratios are $k_{1}=-6.44\%$, $k_{2}=-1.58\%$ and $k_{3}=0.27\%$ for any choice of $\mu_r$. The conventional series is scale-dependent, if setting $\mu_{r}=Q$ to remove the renormalon terms, the ratios are close to the PMC ones, i.e. $k_{1}=-6.25\%$, $k_{2}=-0.93\%$ and $k_{3}=-0.17\%$; However when taking  $\mu_{r}\in[Q/2, 2Q]$, those ratios change greatly, e.g., the uncertainties become $\Delta k_{1}=\left(^{-3.28\%}_{+1.47\%}\right)$, $\Delta k_{2}=\left(^{+10.14\%}_{-2.47\%}\right)$ and $\Delta k_{3}=\left(^{-2.42\%}_{-3.08\%}\right)$, respectively.

According to Eq.~(\ref{equ1}), we obtain $Q_\ast=2.662$~GeV. Because the effective scale $Q_\ast$ is of perturbative nature, its unknown terms shall lead to \textit{the first kind of residual scale dependence}~\cite{Zheng:2013uja}. At present, we don't have too much reliable ways for predict unknown higher order terms. Based on the renormlization scale-indpendent PMC series, the Pad$\acute{e}$ approximation approach (PAA) seems to be realistic. In this process, the PMC scale $Q_*$ is at NLL level as shown in Eq.(13), we need more known terms to do PAA. The PMC scale $Q_*$ is determined by absorbing all the RGE related non-conformal terms into the effective coupling $\alpha_s(Q_*)$. From Eq.(13), the PMC scale is of perturbative and it would suffers from both exponential-suppression and $\alpha_s(Q)$-suppression. Thus, it's reliable to take the NLL-term as missing higher-order term, in other words, as a conservative estimation of \textit{first kind of residual scale dependence}, we take the absolute value of the last known term as the magnitude of the unknown next-to-next-to-leading-log terms ($\rm N^{2}LL$), i.e. $Q_\ast$ up to $\rm N^2LL$-level takes the form

\begin{eqnarray}
\ln\left(\frac{Q^{2}_{\ast}}{Q^2}\right) &=&0.5 -6.669a^{\rm V}_{s}(Q) \pm 6.669a^{{\rm V}}_{s}(Q),
\end{eqnarray}
which leads to a small scale shift $\Delta Q_\ast=\left(^{+0.976}_{-0.714}\right)$ GeV, and hence a small error ($\sim 1\%$) to the ratio $\eta_A$, e.g.
\begin{eqnarray}
\Delta \eta_{A}|_{\rm PMC}=(^{+0.011}_{-0.016}).
\end{eqnarray}

Secondly, it is helpful to have an estimation of the contribution from the uncalculated higher-order terms. In the literature, the Pad$\acute{e}$ approximation approach~\cite{Samuel:1995jc, Samuel:1992qg, Basdevant:1972fe} (PAA) provides an effective do such a prediction. For the PAA, a pQCD approximate $\rho_n$ is expressed as the following $[N/M]$-type form:
\begin{eqnarray}
\rho^{[N/M]}_{n} &=&a^{p}_s\times\frac{b_{0}+b_{1}a_s + \cdots +b_{N}a^{N}_s}{1+c_{1}+\cdots c_{M}a^{M}_s},\\
&=&\sum^{n}_{i=1} C_{i}a^{p+i-1}_{s} +C_{n+1}a^{p+n}_{s} +\cdots,
\end{eqnarray}
where $p$ is the $\alpha_s$-order of the leading-order terms, which is equal to $0$ for the present case of $\eta_A$. The input parameters $b_{i\in[0,N]}$ and $c_{i\in[1,M]}$ can be expressed by using the known perturbative coefficients $C_{i\in[1,n]}$; while the first unknown $(n+1)_{\rm th}$-order coefficient $C_{n+1}$ can be expressed by $b_{i\in[0,N]}$ and $c_{i\in[1,M]}$, and hence by the known coefficients $\{C_{1},...,C_{n}\}$. For the present considered N$^3$LO-level $ \eta_{A}$, since it already shows good convergence for both conventional and PMC series, we take the preferable $[0/n-1]$-type PAA to estimate the contribution of the unknown terms~\cite{Du:2018dma}, which is consistent with the ``Generalized Crewther Relations"~\cite{Shen:2016dnq} and the Gell-Mann-Low method for fixing the coupling constant of quantum electrodynamics~\cite{GellMann:1954fq}. Specifically, the N$^{4}$LO-term for either the conventional series or the PMC series is
\begin{equation}
r^{\rm V}_{4} = \frac{2r^{\rm V}_{1} r^{\rm V}_{2} r^{\rm V}_{3} -r^{{\rm V},3}_{2}}{r^{{\rm V},2}_{1}}
\end{equation}
and
\begin{equation}
{r}^{\rm V}_{4,0} = \frac{2r^{\rm V}_{1,0} r^{\rm V}_{2,0} r^{\rm V}_{3,0}-r^{{\rm V},3}_{2,0}}{r^{{\rm V},2}_{1,0}},
\end{equation}
respectively. Then the predicted N$^4$LO-terms for conventional and PMC series of $\eta_{A}$ are $\eta_{A}|^{\rm N^{4}LO}_{\rm Conv.} = r^{\rm V}_{4} a^{{\rm V},4}_{s}(\mu_{r})$ and $\eta_{A}|^{\rm N^{4}LO}_{\rm PMC} = {r}^{\rm V}_{4,0}a^{{\rm V},4}_{s}(Q_\ast)$, respectively.

\begin{table}[htb]
\begin{tabular}{c c }
\hline
 ~~  ~~ & ~~PAA prediction of $\rm N^{4}LO$-terms~~ \\
\hline
~~~$\eta_{A}|_{\rm Conv.}$~~~   & $-0.0003^{-0.0358}_{-0.0287}$ \\
$\eta_{A}|_{\rm PMC}$    & $+0.0023$ \\
\hline
\end{tabular}
\caption{The $\rm N^{4}LO$-terms of  $\eta_{A}$'s conventional (Conv.) and PMC series predicted by using the [0/2]-type PAA. The prediction of conventional series is for $\mu_{r}\in[Q/{2}, 2Q]$.}
\label{tab3}
\end{table}

We present the PAA predictions of the uncalculated $\eta_{A}$ $\rm N^{4}LO$-terms under PMC and conventional scale-setting approaches in Table~\ref{tab3}, where we have implicitly set $\mu_{r}=Q$ to give the prediction for conventional series. The PAA prediction for conventional series has large scale dependence due to highly scale-dependent renormalon terms, which are proportional to $\beta^{n}_0\ln^{n}(\mu_r^2/Q^2)$, and in effect the PAA amplifies the renormalon divergence. Thus a more accurate prediction can indeed be achieved by using the scale invariant PMC series. The predicted magnitude of the $\rm N^{4}LO$-terms could be treated as the \textit{second kind of residual scale dependence}~\cite{Zheng:2013uja}.

The uncalculated higher-order terms lead to (residual) scale uncertainties for (PMC) conventional series, and the squared average of the above two (residual) scale dependences lead to a total scale error for $\eta_A$, e.g.,

\begin{eqnarray}
\Delta\eta_{A}|^{\rm High~order}_{\rm Conv.} &=&  \left(^{+0.0570}_{-0.0499}\right), \label{CONVHighorder} \\
\Delta\eta_{A}|^{\rm High~order}_{\rm PMC}  &=&   \left(^{+0.0115}_{-0.0166}\right), \label{PMCHighorder}
\end{eqnarray}

which shows that the PMC series has a much smaller scale uncertainty due to uncalculated terms.

Thirdly, except for the scale uncertainty due to uncalculated perturbative terms, there is also uncertainty caused by the error of $\alpha_s$ fixed-point error $\Delta\alpha_s(M_Z)$. By taking $\Delta\alpha_{s}(M_Z)=\pm0.0010$~\cite{Zyla:2020pa}, we obtain
\begin{eqnarray}
\Delta\eta_{A}|^{\Delta\alpha_{s}(M_Z)}_{\rm Conv.} &=& \left(^{+0.0020}_{-0.0021}\right), \\
\Delta\eta_{A}|^{\Delta\alpha_{s}(M_Z)}_{\rm PMC}  &=& \left(^{+0.0024}_{-0.0025}\right).
\end{eqnarray}

For conventional scale-setting approach, the error caused by $\Delta\alpha_{s}(M_Z)$ is about twenty-five times smaller than the error (\ref{CONVHighorder}) caused by the unknown higher-order terms. While for the PMC single-scale setting approach, the error caused by $\Delta\alpha_{s}(M_Z)$ is at the same order of the one (\ref{PMCHighorder}) caused by unknown higher-order terms. Since the PMC uses the RGE to fix the correct $\alpha_s$-running behavior of $\eta_A$, and inversely, its prediction depends heavily on the precise value of $\alpha_s(M_Z)$. This explains why $\Delta\eta_{A}|^{\Delta\alpha_{s}(M_Z)}_{\rm PMC}$ is slightly larger than $\Delta\eta_{A}|^{\Delta\alpha_{s}(M_Z)}_{\rm Conv.}$. Thus more precise measurements on $\alpha_s(M_Z)$ is important for achieving more precise pQCD predictions.

In combination of all the above mentioned errors, we finally obtain
\begin{eqnarray}
\eta_{A}|_{\rm Conv.}&=&0.9265^{+0.0570}_{-0.0499},\\
\eta_{A}|_{\rm PMC}&=&0.9225^{+0.0117}_{-0.0168}.
\end{eqnarray}

\begin{table}[htb]
\centering
\begin{tabular}{c c c c}
\hline
~~~~~~ & ~~PMC~~& ~~Conv.~~  \\
\hline
~~OPAL~partial~reco  \cite{Abbiendi:2000hk}~~  & $42.72^{+1.85}_{-1.94} $ & $42.83^{+3.18}_{-2.92} $ \\
BELLE \cite{Waheed:2018djm}  & $39.94^{+1.18}_{-1.29} $ & $40.04^{+2.69}_{-2.41} $ \\
BABAR~global~fit  \cite{Aubert:2008yv} & $40.67^{+1.21}_{-1.33}$  & $40.77^{+2.74}_{-2.46} $ \\
OPAL~excl  \cite{Abbiendi:2000hk}  & $41.92^{+2.20}_{-2.27}$ & $42.03^{+3.36}_{-3.12} $ \\
BABAR~excl  \cite{Aubert:2007rs} & $39.53^{+1.21}_{-1.31}$ & $39.63^{+2.68}_{-2.40}$ \\
BABAR~$D^{\ast0}$ \cite{Aubert:2007qs}  & $40.89^{+1.38}_{-1.48}$ & $41.00^{+2.83}_{-2.56}$ \\
DELPHI~partial~reco \cite{Abreu:2001ic}  & $40.44^{+1.99}_{-2.06} $ & $40.54^{+3.16}_{-2.92}$ \\
DELPHI~excl  \cite{Abdallah:2004rz}  & $41.33^{+2.29}_{-2.34} $ &  $41.45^{+3.40}_{-3.16}$ \\
\hline
\end{tabular}
\caption{The values of $|V_{\rm cb}|$ ($\times 10^{-3}$) using the PMC and conventional (Conv.) pQCD predictions for $\eta_A$, which are derived by using the data given by various experiments group and under CLN parameterization~\cite{Abbiendi:2000hk, Aubert:2008yv, Aubert:2007rs, Aubert:2007qs, Abreu:2001ic, Abdallah:2004rz, Waheed:2018djm}.}
\label{tab4}
\end{table}

\begin{figure}[htb]
\centering
\includegraphics[width=0.48\textwidth]{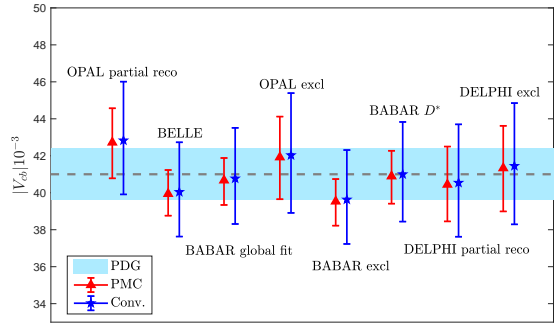}
\caption{The values of $|V_{\rm cb}|$ using the PMC and conventional (Conv.) pQCD predictions for $\eta_A$, where the red and blue error bars represent the PMC and conventional results, respectively. The shaded band is the PDG average value~\cite{Zyla:2020pa}.}
\label{fig2}
\end{figure}

To extract the value of $|V_{\rm cb}|$, we refer to the same parameterization method as the Heavy Flavor Averaging Group(HFLAV) for the parameterization of the form factor function, i.e. the Caprini, Lellouch, and Neubert (CLN)~\cite{Caprini:1997mu} parameterization. Table~\ref{tab4} gives the values of $|V_{\rm cb}|$ ($\times 10^{-3}$) using PMC and conventional (Conv.) pQCD predictions for $\eta_A$, which are derived by using the data given by CLN experimental parameterization~\cite{Abbiendi:2000hk, Waheed:2018djm, Aubert:2008yv, Aubert:2007rs, Aubert:2007qs, Abreu:2001ic, Abdallah:2004rz}. In Table~\ref{tab4}, the errors are calculated by using the following formula,
\begin{eqnarray}
\Delta|{\rm V}_{\rm cb}|&=&\frac{\rm exp}{{\cal{F}}(1)}\sqrt{(\Delta_{\rm exp}/{\rm exp})^2 +(\Delta_{{\cal{F}}(1)}/{\cal{F}}(1))^{2}},
\label{eq29}
\end{eqnarray}
where $\Delta_{\mathcal{F}(1)}$ and $\Delta_{\rm exp}$ represent theoretical and experimental uncertainties, respectively. Table~\ref{tab4} shows that the PMC predictions are more precise than the conventional ones due to much less scale uncertainties. This fact can be more clearly shown by Fig.~\ref{fig2}, in which comparisons of the various predicted values of $|V_{\rm cb}|$ are given. The weighted average of those theoretical values can be calculated by using the method described in detail in Ref.\cite{Zyla:2020pa}, i.e.
\begin{eqnarray}
\overline{x}\pm\delta\overline{x}&=&\frac{\sum_{i}w_{i}x_{i}}{\sum_{i}w_{i}}\pm\left(\sum_{i}w_{i}\right)^{1/2},
\label{eq26}
\end{eqnarray}
where $\overline{x}$ is the central value of the concern parameter (e.g. $|V_{\rm cb}|$) and $\delta\overline{x}$ is its uncertainty. Here $x_{i}$ stands for the $i_{\rm th}$ given value, $\delta_i$ is the uncertainty of the $i_{\rm th}$ value, and $w_{i}=1/(\delta_i x_{i})^{2}$ is the weight factor. We then obtain the weighted average of $|V_{\rm cb}|$,

\begin{eqnarray}
|V_{\rm cb}|_{\rm Conv.} &=& (40.90^{+1.05}_{-1.00})\times10^{-3},  \\
|V_{\rm cb}|_{\rm PMC}  &=& (40.60^{+0.53}_{-0.57})\times10^{-3}.
\end{eqnarray}

Those two values are consistent with the PDG average value $|V_{\rm cb}|_{\rm PDG}= (40.8\pm 1.4)\times10^{-3}$ within errors~\cite{Workman:2022ynf}. As a comparison, the recent value given by the HFLAV is $|V_{\rm cb}|_{\rm HFLAV}=(38.76\pm0.55)\times10^{-3}$~\cite{HFLAV:2019otj}, which deviates from the PDG average value by $0.164\sigma$, the present PMC prediction by $1.309\sigma$ and the present conventional prediction by $1.072\sigma$, respectively.

As a summary, in this paper, we have given a detailed analysis on the perturbative nature of the parameter $\eta_{A}$ for the $B\to D^{\ast}\ell\bar{\nu}$ process up to $\rm N^{3}LO$-level.

By applying the PMC single-scale setting approach, the conventional scale ambiguity is removed. Thus the PMC scale-invariant series provides a better platform for achieving a more precise fixed-order pQCD prediction. To compare with a larger uncertainty caused by unknown terms, $\Delta\eta_{A}|^{\rm High~order}_{\rm Conv.} =  \left(^{+0.0570}_{-0.0499}\right)$ for conventional series, the PMC series has a much smaller error, $\Delta\eta_{A}|^{\rm High~order}_{\rm PMC}=\left(^{+0.0117}_{-0.0168}\right)$. Then a more accurate prediction on $|V_{\rm cb}|$ can be achieved, which is consistent with the PDG average within errors. Thus our present results emphasize the necessity of a proper renormalization scale-setting approach during the pQCD calculation. \\

\noindent {\bf Acknowledgments:} This work was supported by the graduate research and innovation foundation of Chongqing, china (No.CYB21045 and No.ydstd1912), by the Natural Science Foundation of China under Grant No.12005028, No.12175025 and No.12147102, and by the China Postdoctoral Science Foundation under Grant No.2021M693743, the authors Hua Zhou and Qing Yu thank the financial support from the China Scholarship Council.

\end{document}